\documentclass[twocolumn,prd,superscriptaddress]{revtex4-1}
\usepackage[colorlinks]{hyperref}
\usepackage{graphicx}
\usepackage{amsfonts}
\usepackage{amssymb}
\usepackage{amsbsy}
\usepackage{amsmath}
\usepackage{latexsym}
\usepackage{color}

\DeclareFontFamily{OT1}{rsfs}{} 
\DeclareFontShape{OT1}{rsfs}{m}{n}{<-7> rsfs5 
    <7-10> rsfs7 <10-> rsfs10}{}   
\DeclareMathAlphabet{\scr}{OT1}{rsfs}{m}{n}

\newcommand*{\di}{\partial}

\def\be {\begin{equation}}
\def\ee  {\end{equation}}
\def\bea {\begin{eqnarray}}
\def\eea {\end{eqnarray}}
\def\nn {\nonumber}

\renewcommand*{\k}{\hat{k}}

\newcommand*{\M}{M_{\star}}

\def\ce{\mathrm{ce}}
\def\se{\mathrm{se}}

\def\k{\mathbf{k}}
\def\x{\mathbf{x}}
\def\y{\mathbf{y}}

\begin{document}

\title{High energy modifications of blackbody radiation and dimensional reduction}

\author{Viqar Husain}
\email{husain@unb.ca}
\affiliation{Department of Mathematics and Statistics, University of New Brunswick, Fredericton, NB, Canada E3B 5A3} 
 
\author{Sanjeev S.\ Seahra}
\email{sseahra@unb.ca} 
\affiliation{Department of Mathematics and Statistics, University of New Brunswick, Fredericton, NB, Canada E3B 5A3} 

\author{Eric J.\ Webster}
\email{ewebster@uwaterloo.ca}
\affiliation{Department of Applied Mathematics, University of Waterloo, Waterloo, Ontario, N2L 3G1, Canada}

\pacs{04.60.Ds}

\begin{abstract}

Quantization prescriptions that realize generalized  uncertainty relations  (GUP)  are motivated by quantum gravity arguments  that incorporate a fundamental length scale.   We apply two such methods,  polymer and deformed Heisenberg quantization,  to scalar field theory  in Fourier space. These alternative quantizations  modify the oscillator spectrum for each mode, which in turn affects the blackbody distribution.  We find that for a large class of modifications, the equation of state relating pressure $P$ and energy density $\rho$ interpolates between $P=\rho/3$ at low $T$ and $P=2\rho/3$ at high $T$, where $T$ is the temperature.  Furthermore, the Stefan-Boltzman law gets modified from $\rho \propto T^{4}$ to $\rho \propto T^{5/2}$ at high temperature.  This suggests an effective reduction to 2.5 spacetime dimensions at high energy. 

\end{abstract}

\maketitle

\section{Introduction}

The inherent difficulties in constructing a quantum theory of gravity have led many to believe that conventional physical theories need to be significantly altered on very small scales.  For example, string theory predicts extra dimensions of exceedingly small size, while loop quantum gravity posits that standard Schr\"odinger quantization is merely an approximation to polymer quantization valid on super-Planckian scales.  A naturally interesting question is how exotic small scale physics might affect quantum field theory (QFT).  A number of authors have looked at this issue in the context of Hawking radiation \cite{Unruh:1994je,Corley:1996ar,Berger:2011wd}, the two-point function and wave propagation of a free scalar field \cite{Hossain:2009vd,Hossain:2010eb,Husain:2013ul}, and the trans-Planckian problem for the generation of primordial perturbations during inflation \cite{Brandenberger:2000wr,Chu:2000ww,Easther:2001fi,Kempf:2000ac,Kempf:2001fa,Martin:2000xs,Niemeyer:2000eh,Niemeyer:2001qe,Starobinsky:2001kn,Danielsson:2002kx,Lizzi:2002ib,Niemeyer:2002kh,Bozza:2003pr,Burgess:2002ub,Burgess:2003zw,Hassan:2002qk,Martin:2003kp,Shankaranarayanan:2002ax,Ashoorioon:2004vm,Brandenberger:2004kx,Shankaranarayanan:2005cs,Kempf:2006wp,Brandenberger:2009rs,Piao:2009ax,Seahra:2012un}.  

In this work we are interested in how exotic high energy physics may impact the behaviour of blackbody radiation at high temperature.  This problem has been considered before:  Effects of the deformation of the quantum commutator algebra between position and momentum on a scalar gas have been investigated in the classical limit \cite{Chang:2002rr}, and by employing modified field quantization in Fourier space \cite{Mania:2011qy}.  Possible ramifications of spacetime non-commutivity on the blackbody spectrum were discussed in \cite{Fatollahi:2006rt}.  The dependence of the thermal characteristics of the gas on the existence of non-compact \cite{Alnes:2007mz} and compact \cite{Ramos:2011gf} extra dimensions has been reported.  Finally in \cite{Nozari:2012yq}, the combined effects of modified dispersion relations and extra dimensions were studied.

 We restrict our attention to high-energy effects that modify the energy levels of field quanta.  It is well-known that if one considers the Fourier transform of a non-interacting scalar field, the Hamiltonian governing the evolution of the Fourier coefficients is that of a collection of decoupled one-dimensional simple harmonic oscillators (SHOs).   All the familiar results concerning the quantum behaviour of a free scalar field can be derived by quantizing these oscillators using ordinary quantum mechanics \cite{Hossain:2010eb,Husain:2013ul}.  Hence one of the most straightforward ways to incorporate high-energy modifications into QFT is to apply non-Schr\"odinger quantization techniques to these oscillators, as is done in Refs.\ \cite{Berger:2011wd,Hossain:2010eb,Husain:2013ul,Mania:2011qy,Seahra:2012un}.  
 
 The most important consequence for statistical mechanics is that these alternate quantization schemes will change the energy spectrum of each Fourier mode.  That is, the energy level spacing is no longer $\hbar k$, but rather a more complicated function.  This information can be directly used to calculation the partition function of the scalar gas at finite temperature, which in turn specifies all its thermal properties \footnote{For the remainder of this paper, we work in units where $\hbar=c=1$}.

After reviewing the oscillator description of QFT in \S\ref{sec:QFT}, we investigate the implications of   generic modifications to the oscillator energy levels in \S\ref{sec:stat mech} assuming that   that for large wavenumbers,  energy eigenvalues scale like $k^{\alpha}$ rather than $k$.   This implies that the internal energy $\mathcal{U}$ of the gas is $\propto T^{1+3/\alpha}$ and its equation of state is $P = \frac{\alpha}{3} \rho$, where $P$ is the pressure and $\rho$ is the density.

In \S\ref{sec:modifications}, we consider several specific models that result in modifications of the oscillator energy levels.  One class of such models involves modifying the commutator of the field amplitude $\hat\phi_{\k}$ and the field momentum $\hat\pi_{\k}$ in Fourier space as follows:
\begin{equation}
	[\hat\phi_{\k},\hat\pi_{\k}] = i \quad \mapsto \quad [\hat\phi_{\k},\hat\pi_{\k}] = if(\hat\pi_{\k}/\M),
\end{equation}
where $\M$ is an energy scale that indicates the threshold for exotic physics.  Finding the energy levels of the Fourier modes is equivalent to finding the energy levels of a SHO quantized according to the commutator $[\hat{x},\hat{p}]=if(\beta\hat{p})$, where $\beta$ is a dimensional constant.  The SHO energy levels for the $f(x)=1+x^{2}$ case were found in \cite{Kempf:1994su}, while the $f(x)=1-x^{2}$ case was considered in \cite{Husain:2013ul}.  The leading order low temperature corrections to blackbody radiation due to the choice $f=1+x^{2}$ were reported in \cite{Mania:2011qy}.  (In this paper, we do not restrict ourselves to the low temperature regime.)  In \S\ref{sec:GUP}, we give details of the energy level calculations for the choices $f(x)=1\pm x$ and summarize the results in Table \ref{tab:eigenfunctions}.

We then explore a different class of new physics suggested by the background independent (or ``polymer'') approach to quantization that is deployed in loop quantum gravity (LQG) \cite{Thiemann:2007zz}.  The principal difference between this approach and traditional quantization is that  operators corresponding to the momentum do not exist due to the unconventional choice of Hilbert space, but  operators corresponding to the exponentiated momentum are well defined.  The novel features of this approach to quantization have been considered in a number of   quantum mechanical systems \cite{Ashtekar:2002sn,Fredenhagen:2006wp,Corichi:2007tf,Kunstatter:2008qx,Kunstatter:2009ua,Kunstatter:2010fa,Hossain:2010eb,Kunstatter:2012ra}, quantum cosmology \cite{Bojowald:2001xe,Ashtekar:2003hd,lrr-2005-11,Ashtekar:2006rx,Hossain:2009ru}, and quantum field theory \cite{Ashtekar:2002vh,Laddha:2010hp,Hossain:2009vd,Husain:2010gb,Hossain:2010eb}.  

The statistical mechanics of a polymer quantized harmonic oscillator has also been discussed  \cite{ChaconAcosta:2011vv}, but our results go further.  In \S\ref{sec:polymer}, we review the basic properties of a polymer oscillator and present its energy eigenvalues \cite{Ashtekar:2002sn,Hossain:2010eb}.  An important point is that the polymer quantization scheme for this system carries with it an ambiguity related to the choice of a super-selected sector of the Hilbert space (this issue is discussed in detail in Refs.~\cite{Ashtekar:2002sn,Seahra:2012un}).  Hence, we obtain several possible energy spectra based on how this ambiguity is resolved.

In \S\ref{sec:numeric} we numerically calculate the thermal properties of a scalar gas using the modified energy levels presented in \S\ref{sec:modifications}.  We find that some of the modifications lead to UV divergent integrals for thermodynamic quantities.  All of the other modifications discussed share the property that the energy eigenvalues scale as $k^{2}$ for large $k$, which by the arguments of \S\ref{sec:stat mech}  lead to the high temperature Stefan-Boltzman law $\mathcal{U} \propto T^{5/2}$ and equation of state $P = 2\rho/3$.  We confirm these expectations numerically. We summarize and discuss our results in \S\ref{sec:discussion}.

\section{Oscillator description of quantum field theory}\label{sec:QFT}

We consider the quantum field theory of a scalar field in Minkowski space time.  The Hamiltonian is
\begin{equation}\label{eq:standard H}
H_{\phi} = \int d^{3} x \, \left[ \frac{1}{2} \pi^{2} + \frac{1}{2} (\mathbf{\nabla}\phi)^{2} \right].
\end{equation}
Here, $\pi$ is the momentum conjugate to $\phi$ such that $\{\phi(t,\x),\pi(t,\y)\} = \delta^{(3)}(\x-\y)$.  We decompose $\phi$ and $\pi$ into Fourier modes as follows:
 \begin{align}
\phi(t,\x) &=  \frac{1}{\sqrt{V}}\sum_{\k}{\phi}_{\k}(t) e^{i {\k}\cdot{\x}}, \nn\\
{\phi}_\k(t) &= \frac{1}{\sqrt{V}}\int d^3x\ e^{-i\k\cdot \x} \phi(t,\x),
\end{align}
with a similar expansion for $\pi(t,\x)$; $V$ is the fiducial volume used in our box normalization
\begin{equation}
V = \int d^{3}x.
\end{equation}
After  a suitable redefinition of the independent modes to enforce that $\phi$ is real, the Hamiltonian is 
\begin{equation}\label{eq:Fourier H}
H_{\phi} =  \sum_{\k} H_{\k} = \sum_{\k} \left[ \frac{\pi_{\k}^2}{2} +
\frac{k^{2}}{2} \phi_{\k}^2 \right],
\end{equation}
with the Poisson bracket $\{\phi_\k, \pi_{\k'}\} = \delta_{\k,\k'}$. 

We now quantize the classical system (\ref{eq:Fourier H}).  The structure of the Hamiltonian is that of a collection of decoupled harmonic oscillators labelled by $\k$.  Hence, the quantum state of the system will be
\begin{equation}
	|\psi \rangle = \bigotimes_{\k} |\psi_{\k} \rangle,
\end{equation}
where each of the $|\psi_{\k} \rangle$ satisfy the Schr\"odinger equation
\begin{equation}\label{eq:oscillator Hamiltonian}
	i \di_{t} |\psi_{\k} \rangle = \hat{H}_{\k}  |\psi_{\k} \rangle, \quad \hat{H}_{\k} = \frac{1}{2} \hat{\pi}_{\k}^{2} + \frac{k^{2}}{2} \hat{\phi}_{\k}^{2}.
\end{equation}
 Since the energy eigenstates $ | n_{\k} \rangle$ of $\hat{H}_{\k}$ form a basis, we can express $|\psi_{\k}\rangle$ as 
\begin{equation}
	|\psi_{\k} \rangle(t)  = \sum_{n=0}^{\infty} e^{-iE_{n,k}t} | n_{\k} \rangle, 
\end{equation}
where
\begin{equation}\label{eq:eigenvalue problem}
	 \hat{H}_{\k}  | n_{\k} \rangle = E_{n,k} | n_{\k} \rangle.
\end{equation}
In standard Schr\"odinger quantization, the solution to the eigenvalue problem (\ref{eq:eigenvalue problem}) is the well-known energy eigenstates of the simple harmonic oscillator with $E_{n} = k (n+1/2)$.  

For alternative quantizations, this spectrum is modified for modes with $k \gtrsim \M$, where $\M$ is an energy scale associated with quantum gravity.  To parametrize such effects, let us introduce a dimensional quantity 
\begin{equation}
	g = \frac{k}{\M},
\end{equation}
such that the oscillator spectrum takes the scaling form
\begin{equation}
	\frac{E_{n,k}}{\M} = \varepsilon_{n}(g),
\end{equation}
where the functions $\varepsilon_{n}(g)$ will depend of the specific class of high-energy modification being considered.  The correct low energy physics is recovered if 
\begin{equation}
	\varepsilon_{n}(g) \approx g(n+1/2), \quad g \ll 1.
\end{equation}

\section{Statistical mechanics}\label{sec:stat mech}

\subsection{Formalism}

Let us now turn our attention to the statistical properties of the field at a finite temperature $T$.  The partition function for a given $\k$ mode is
\begin{equation}\label{eq:normal partition}
	Z_{\k}(\beta) = \sum_{n=0}^{\infty} e^{-\beta(E_{n,k}-E_{0,k})},
\end{equation}
where
\begin{equation}
	\beta = \frac{1}{k_\text{B} T},
\end{equation}
and $k_{\text{B}}$ is Boltzmann's constant.  The average energy in the mode (above the ground state) is then
\begin{equation}\label{eq:normal energy}
	\bar{E}_{\k}(\beta)  = -\frac{\di}{\di \beta} \ln Z_{\k}(\beta).
\end{equation}
In three spatial dimensions, the total internal energy $\mathcal{U}$ of the gas is
\begin{equation}\label{eq:normal total energy}
	\mathcal{U} = \frac{V}{\pi^{2}} \int_{0}^{\infty} dk \, k^{2} \bar{E}_{\k}(\beta).
\end{equation}

These formulae may be converted to dimensionless units by introducing
\begin{equation}
	\tilde\beta = \beta \M.
\end{equation}
Then  the energy is given by
\begin{equation}\label{eq:rho}
	\mathcal{U} = V\M^{4} \int_{0}^{\infty} dg \, I(g,\tilde\beta),
\end{equation}
where the dimensionless intensity is 
\begin{equation}\label{eq:intensity}
	I(g,\tilde\beta) = -\frac{g^{2}}{\pi^{2}} \frac{\di}{\di \tilde\beta} \ln \left( \sum_{n=0}^{\infty} e^{-\tilde\beta \Delta\varepsilon_{n}(g) } \right),
\end{equation}
and
\begin{equation}
	\Delta\varepsilon_{n}(g) = \varepsilon_{n}(g) - \varepsilon_{0}(g).
\end{equation}

The  entropy $S$ and pressure $P$ of the gas is obtained from $\mathcal{U}$  in the standard way.  Define the energy and entropy densities by
\begin{equation}
	\rho(T) = \frac{\mathcal{U}}{V}, \quad s(T) = \frac{S}{V}.
\end{equation}
Then the thermodynamic identities
\begin{equation}
	\left( \frac{\di S}{\di \mathcal{U}} \right)_{V} = \frac{1}{T}, \quad \left( \frac{\di S}{\di T} \right)_{V} = \left( \frac{\di S}{\di \mathcal{U}} \right)_{V} \! \left( \frac{\di \mathcal{U}}{\di T} \right)_{V},
\end{equation}
yield the following formula for the entropy density
\begin{equation}\label{eq:entropy}
	s(T) = \frac{\rho(T)}{T} + \int_{0}^{T} \frac{\rho(\theta)}{\theta^{2}} d\theta,
\end{equation}
where we have assumed
\begin{equation}
	s(0) = 0, \quad \lim_{T\rightarrow 0^{+}} \frac{\rho}{T} = 0.
\end{equation}
Using the Maxwell relation
\begin{equation}
	\left( \frac{\di S}{\di V} \right)_{T} = \left( \frac{\di P}{\di T} \right)_{V}, 
\end{equation}
and noting that the pressure is intensive, we obtain
\begin{equation}\label{eq:pressure}
	P(T) = \int_{0}^{T} s(\theta) d\theta.
\end{equation}

To summarize, the formulae in this subsection give the energy $\mathcal{U}$, entropy $S$ and pressure $P$ of an scalar gas in terms of the temperature $T$ and volume $V$, 
using the energy eigenvalues $\varepsilon_{n}(g)$ of modified oscillators corresponding to different $\k$ modes of the scalar field. All other thermodynamic quantities  are similarly determined.

\subsection{Low and high temperature limits}\label{sec:low and high T}

In the low temperature limit $\tilde\beta \gg 1$, the exponential functions appearing in (\ref{eq:intensity}) are  highly damped for $g \gtrsim 1$; hence, we can use low-$g$ approximation for $\Delta\varepsilon_{n}$ . Then the correct  low energy limit 
\begin{equation}
	\Delta \varepsilon_{n} \approx ng, \quad g \ll 1.
\end{equation}
gives the usual blackbody spectrum at low temperature, 
\begin{equation}
	I = \frac{1}{\pi^{2}} \frac{g^{3}}{e^{\tilde\beta g}-1} = \frac{1}{\M^{3}\pi^{2}} \frac{k^{3}}{e^{\beta k} -1}, 
\end{equation}
and Stefan-Boltzman law 
\begin{equation}
	\mathcal{U} = \frac{\pi^{2}}{15} V k^{4}_{\text{B}}T^{4}.
\end{equation}
The expressions (\ref{eq:entropy}) and (\ref{eq:pressure}) then yield familiar results 
\begin{equation}
	s = \frac{4}{3} \frac{\rho}{T}, \quad P = \frac{1}{3} \rho.
\end{equation}

In the high-temperature limit $\tilde\beta \ll 1$, we cannot write down a closed form expression for the intensity without knowing the energies $\varepsilon_{n}(g)$.  However, we would expect the integrand in (\ref{eq:rho}) to be peaked at some high wavenumber $g \gg 1$.  Furthermore, it is reasonable to assume that the energy differences scale as some power of $g$ in this limit:
\begin{equation}
	\Delta\varepsilon_{n} \approx g^{\alpha} f_{n}, \quad g \gg 1,
\end{equation}
where the numbers $f_{n}$ are independent of $g$.  Under such an assumption, we find
\begin{equation}
	\mathcal{U} = K V T^{\frac{3}{\alpha}+1}, 
\end{equation} 
where 
\begin{equation}
\quad K =  \kappa k_\text{B}^{\frac{3}{\alpha}+1} M_{\star}^{3-\frac{3}{\alpha}}, 
\end{equation}
and 
\begin{equation}
	\kappa = \frac{1}{\alpha\pi^{2}} \int_{0}^{\infty} dx \, x^{3/\alpha} \left( - \frac{\di}{\di x} \ln \sum_{n=0}^{\infty} e^{-f_{n}x} \right),
\end{equation}
is a dimensionless constant  which depends on the precise form of $f_{n}$.  At high temperatures, the integrals in  (\ref{eq:entropy}) and (\ref{eq:pressure}) are dominated by the high $\tilde{T}$ behaviour of $\rho(\tilde T)$, which give  
\begin{equation}
	s = \left(1+\frac{\alpha}{3} \right) \frac{\rho}{T} , \quad P = \frac{\alpha}{3} \rho.
\end{equation}
In other words, the equation of state is $P/\rho = \alpha/3$ at high temperature. A similar analysis was carried out in \cite{Das-Husain} in the context of holography.

\section{Alternative quantizations of the oscillator}\label{sec:modifications}

In the previous section, we saw how the oscillator energy spectrum  (\ref{eq:oscillator Hamiltonian}) fixes the thermal properties of the scalar gas.  In this section we will see how the this spectrum is  modified by GUP and polymer quantizations as a prelude to deriving its effects on the blackbody distribution. 

\subsection{GUP quantization}\label{sec:GUP}

Several approaches to quantum gravity suggest that one should modify the fundamental commutator $[\hat{x},\hat{p}]=i$ of quantum mechanics at high energies, which leads to deformed uncertainty relations.  In terms of the QFT oscillator variables of \S\ref{sec:QFT}, we write these as
\begin{equation}
	[\hat{\phi}_{\k} ,\hat{\pi}_{\k} ] = i  f\left( \frac{\hat{\pi}_{\k}}{M_{\star}^{1/2}} \right),
\end{equation}
where $f(0) = 1$ in order to recover the standard commutator for $\pi_{\k} \ll M_{\star}^{1/2}$.  We introduce a basis of momentum eigenstates $|\pi_{\k}\rangle$ and express the Hamiltonian eigenstates $|n_{\k}\rangle$ as wavefunctions:
\begin{equation}\label{eq:momentum wavefunction}
	\langle \pi_{\k} | n_{\k} \rangle = \psi_{n}(\pi_{\k}).
\end{equation}
In order to faithfully reproduce the above commutator, we take the action of $\hat{\phi}_{\k}$ and $\hat{\pi}_{\k}$ on these wavefunctions as:
\begin{subequations}
\begin{align}
	\langle \pi_{\k} | \hat\phi_{\k} | n_{\k} \rangle &  = i f (\pi_{\k}/M_{\star}^{1/2}) {\di_{\pi_{\k}}} \psi_{n}(\pi_{\k}), \\ \langle \pi_{\k} | \hat{\pi}_\k |n_{\k} \rangle & = \pi_{\k }\psi_{n}(\pi_{\k}).
\end{align}
\end{subequations}
In this representation the eigenvalue equation (\ref{eq:eigenvalue problem}) reads 
\begin{equation}
	E_{n,k} \psi_{n}(\pi_{k}) = \left\{ \frac{\pi_{\k}^{2}}{2} - \frac{k^{2}}{2} \left[ f \left( \frac{\hat{\pi}_{\k}}{M_{\star}^{1/2}} \right) \frac{\di}{\di\pi_{\k}} \right]^{2} \right\} \psi_{n}(\pi_{k}).
\end{equation}
Let us change variables according to
\begin{equation}
	\pi_{\k} = M_{\star}^{1/2} P(z), \quad z =  \int_{0}^{P(z)} \frac{du}{f(u)},
\end{equation}
and define the following dimensionless quantities 
\begin{equation}\label{eq:case A defs}
	g = \frac{k}{\M}, \quad \Psi_{n}(z)= \psi_{n}(P(z)), \quad \kappa_{n} = \frac{E_{n,k}}{g^{2}\M}.
\end{equation}
Then the eigenvalue equation reads 
\begin{equation}\label{eq:case A Schrodinger}
	\kappa_{n} \Psi_{n}(z) = \left[ -\frac{1}{2}\frac{\di^{2}}{\di z^{2}} + V(z) \right] \Psi_{n}(z),
\end{equation}
where the potential is given by
\begin{equation}
V(z) = \frac{P^{2}(z)}{2g^{2}}.
\end{equation}

\subsubsection*{Case A0: $f(x) = 1$}

This is the standard result that follows from the  Heisenberg algebra 
\begin{equation}
	[\hat{\phi}_{\k} ,\hat{\pi}_{k} ] = i.
\end{equation}
It leads to the eigenvalue equation
\begin{equation}
	\kappa_{n} \Psi_{n}(z) = \left[ -\frac{1}{2}\frac{\di^{2}}{\di z^{2}} + \frac{z^{2}}{2g^{2}} \right] \Psi_{n}(z),
\end{equation}
where $\displaystyle \pi_{\k} = M_{\star}^{1/2} z.$
Its solutions are given in terms of Hermite polynomials and are summarized in Table \ref{tab:eigenfunctions}.

\subsubsection*{Case AI: $f(x) = 1+x^{2}$}

For the modified commutator  
\begin{equation}
	[\hat{\phi}_{\k} ,\hat{\pi}_{k} ] = i  \left(1 +  \frac{\hat{\pi}_{\k}^{2}}{M_{\star}} \right),
\end{equation}
 the eigenvalue equation is 
\begin{equation}
	\kappa_{n} \Psi_{n}(z) = \left[ -\frac{1}{2}\frac{\di^{2}}{\di z^{2}} + \frac{\tan^{2}(z)}{2g^{2}} \right] \Psi_{n}(z),
\end{equation}
where  $ \displaystyle \pi_{\k} = M_{\star}^{1/2} \tan z.$
	
\begin{figure*}
	\includegraphics[width=\textwidth]{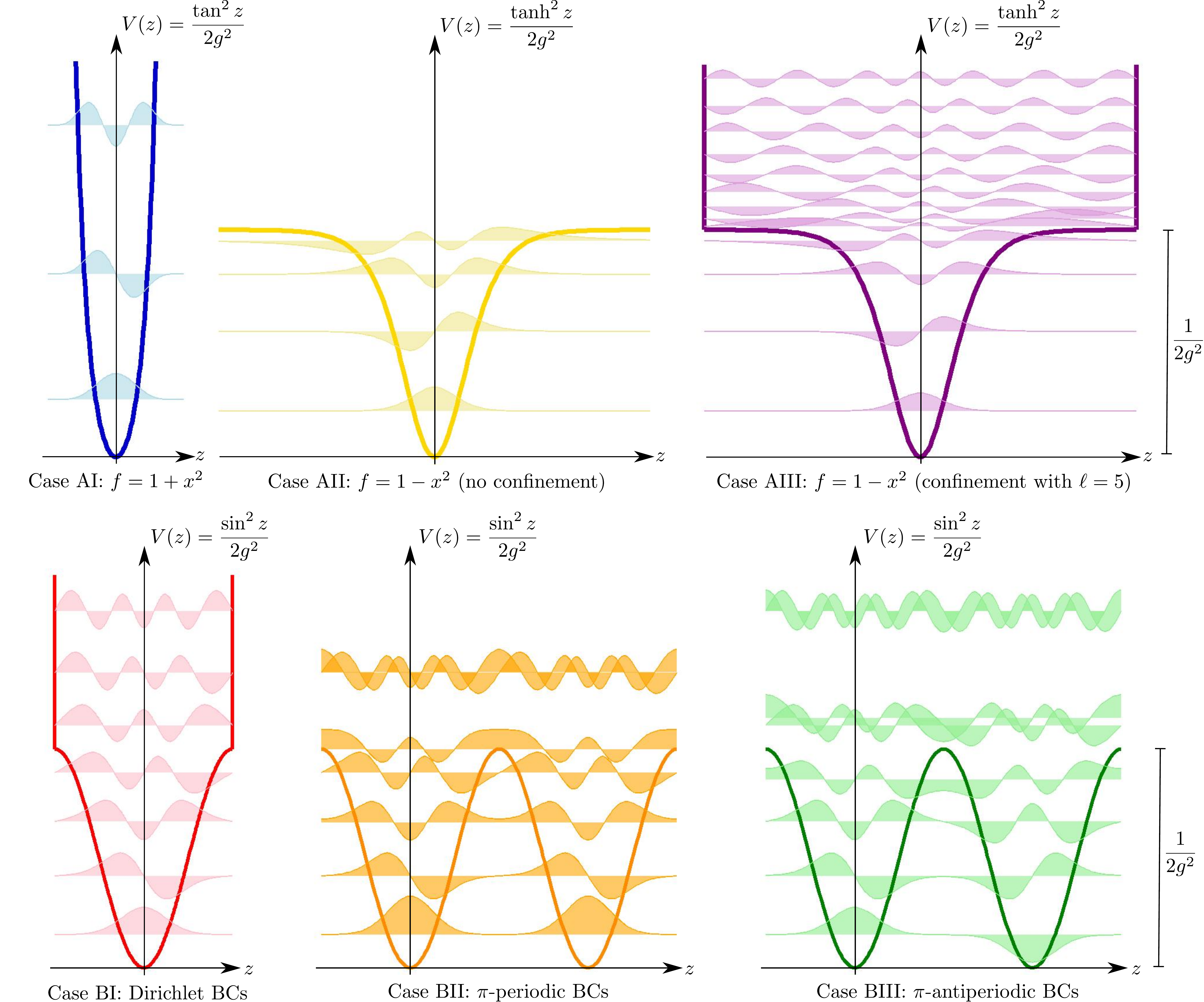}
	\caption{Potentials and energy eigenfunctions in the polymer quantization case.  We have take $g = 0.15$ for Cases AI--AIII and $g=0.225$ for Cases BI--BIII.  Also, we take $\ell = 5$ for Case AIII.}\label{fig:eigenfunctions}
\end{figure*}

\begin{figure*}
	\includegraphics[width=0.9\textwidth]{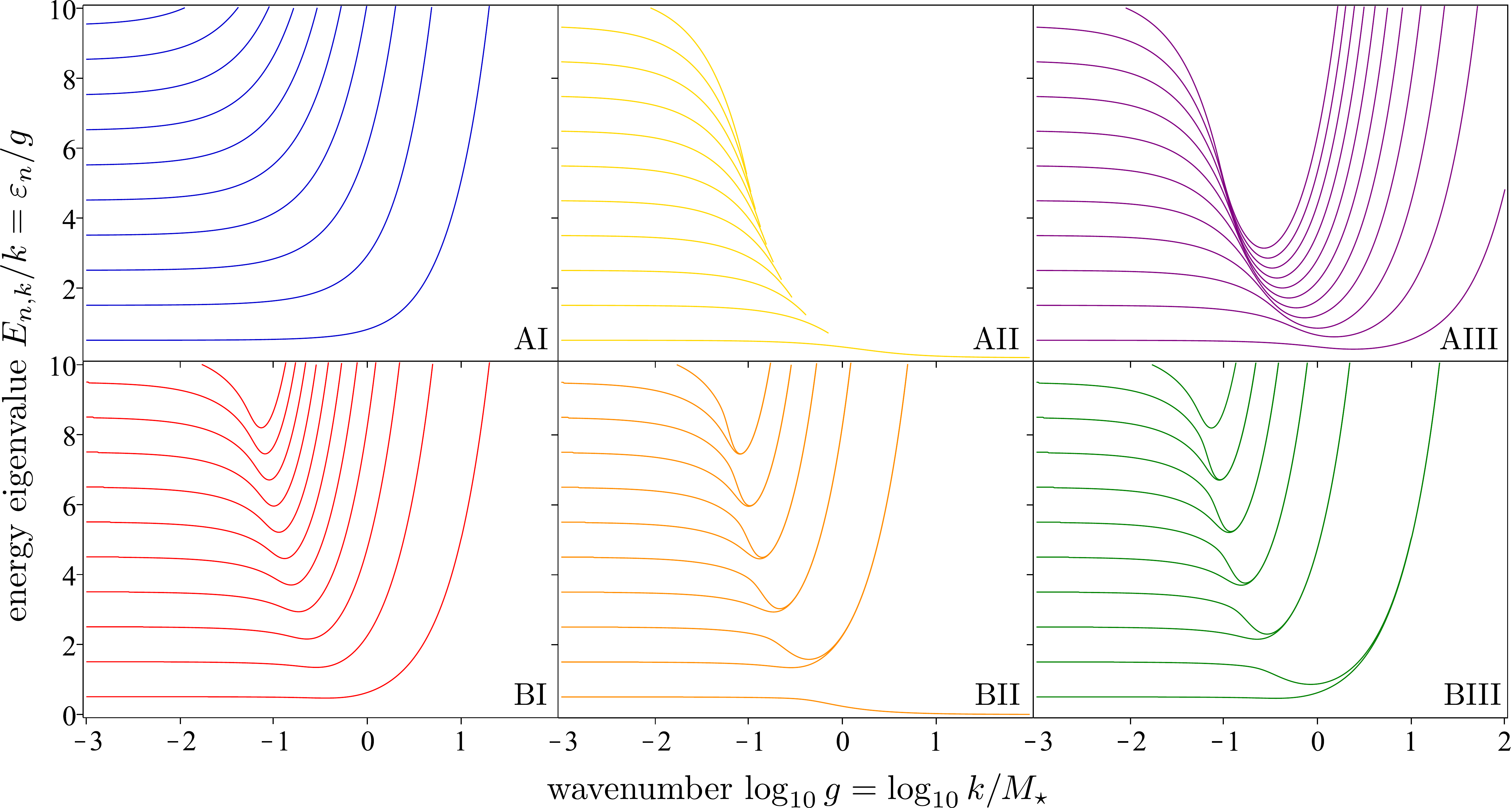}
	\caption{Energy eigenvalue spectra for the various modified field theories considered.  For case AIII, we take $\ell = 5$.}\label{fig:eigenvalues}
\end{figure*}

This eigenvalue problem is analytically solvable in terms of ultraspherical (Gegenbauer) polynomials \cite{abramowitz+stegun} for eigenfunctions satisfying Dirichlet boundary conditions:
\begin{equation}
	\Psi_{n}(\pm \pi/2) = 0,
\end{equation}
and inner product 
\begin{equation}
(\Psi_{n},\Psi_{m})_{\pi/2} \equiv \int_{-\pi/2}^{\pi/2} \Psi_{n}^*(z)\Psi_{m}(z) dz = \delta_{m,n}.
\end{equation}
(Note that since the potential has second order poles at $z = \pm \pi/2$, it is possible to choose different self-adjoint extensions of the Hamiltonian---i.e.~alternative boundary conditions---for certain values of $g$, see Ref. \cite{Narnhofer1974} for details of such potentials.)  Explicit formulae for the energy eigenfunctions and eigenvalues are given in Table \ref{tab:eigenfunctions}, and they are plotted in Figures \ref{fig:eigenfunctions} and \ref{fig:eigenvalues}, respectively.  The eigenvalues have the  asymptotic behaviour 
\begin{equation}
	\varepsilon_{n}(g) \approx \begin{cases} g(n+\tfrac{1}{2}), & g \ll 1, \\ \frac{1}{2} g^{2} (n+1)^{2}, & g \gg 1. \end{cases}
\end{equation}
It is therefore evident that  the usual result at low $g$ is recovered.  At high $g$, the energy differences are
\begin{equation}
\Delta \varepsilon_{n}(g) \approx \tfrac{1}{2} g^{2} n(n+2)  = g^{2} f_{n}.
\end{equation}

\begin{table*}

\caption{The six classes Schr\"odinger equation potentials, boundary conditions and solutions leading to the modified to the oscillator spectra considered in this paper, as well as the standard case A0}\label{tab:eigenfunctions}

\begin{ruledtabular}

{
\begin{tabular}{lccccc}

 \centering{Case} & potential $V(z)$ &  boundary conditions & energy eigenfunctions $\Psi_{n}(z)$ & energy eigenvalues $\varepsilon_{n} = g^{2}\kappa_{n}$  \\  \hline

A0\footnotemark[1] & $\displaystyle \frac{z^{2}}{2g^{2}}$ & $\begin{array}{c} \text{Dirichlet} \\ \displaystyle \lim_{z\rightarrow \pm \infty}\Psi_{n}(z)=0 \end{array}$  & $\displaystyle \frac{1}{(\pi g)^{1/4} (2^{n} n!)^{1/2} } \exp\left(-\frac{z^{2}}{2g} \right) H_{n}\left( \frac{z}{\sqrt{g}} \right) $  & $\displaystyle  g \left(n + \frac{1}{2}\right)$

\\ [4mm] 

AI\footnotemark[2]\textsuperscript{,}\footnotemark[3]\textsuperscript{,}\footnotemark[4] & $\displaystyle \frac{\tan^{2}z}{2g^{2}}$ & $\begin{array}{c} \text{Dirichlet} \\ \Psi_{n}(-\pi/2)= \Psi_{n}(\pi/2)=0 \end{array}$  & $L_{n,l} (\cos z)^{n+l} C_{n}^{(-n-l+\frac{1}{2})}\left( i\tan z \right)$ & $\displaystyle\frac{1}{2} g^{2} (n^{2} + 2nl + l)$

\\ [4mm] 

AII\footnotemark[2]\textsuperscript{,}\footnotemark[3]\textsuperscript{,}\footnotemark[5]  & $\displaystyle \frac{\tanh^{2}z}{2g^{2}}$ &  $\begin{array}{c} \text{Dirichlet} \\ \displaystyle \lim_{z\rightarrow \pm \infty}\Psi_{n}(z)=0 \end{array}$ & $M_{n,j} (\text{sech}\, z )^{j-n} C_{n}^{(-n+j+\frac{1}{2})}\left(\tanh z \right)$ & $\displaystyle\frac{1}{2} g^{2} (-n^{2} + 2nj + j)$

\\ [4mm] 

AIII\footnotemark[3]\textsuperscript{,}\footnotemark[6] & $\displaystyle \frac{\tanh^{2}z}{2g^{2}}$ & $\begin{array}{c} \text{Dirichlet} \\ \Psi_{n}(-\ell)= \Psi_{n}(\ell)=0 \end{array}$  & $N^{(+)}_{n,j} P_{j}^{i\nu_{n}}(\tanh z) + N^{(-)}_{n,j} P_{j}^{i\nu_{n}}(-\tanh z)$ & $\displaystyle\frac{1}{2} (g^{2} \nu_{n} + 1)$

\\ [4mm] 

BI\footnotemark[7] & $\displaystyle \frac{\sin^{2}z}{2g^{2}}$ & $\begin{array}{c} \text{Dirichlet} \\ \Psi_{n}(-\pi/2)= \Psi_{n}(\pi/2)=0 \end{array}$  & $\displaystyle \frac{1}{\sqrt{\pi}}\mathrm{se}_{n+1}\left(\zeta,z+\pi/2 \right)$ & $\displaystyle \frac{g}{4} \left[ {2g B_{n+1}\left( \zeta \right)+1/g} \right]$ 

\\ [4mm]

BII\footnotemark[7] & $\displaystyle \frac{\sin^{2}z}{2g^{2}}$ & $\begin{array}{c} \text{$\pi$-periodic} \\ \Psi_{n}(-\pi/2)= \Psi_{n}(\pi/2) \end{array}$  & $\displaystyle \frac{1}{\sqrt{\pi}}\begin{cases}
	    \text{ce}_{n}(\zeta,z+\pi/2 ), & n \text{ even}  \\
	    \text{se}_{n+1}(\zeta,z+\pi/2 ), & n \text{ odd} 
	\end{cases}$ & $\displaystyle \frac{g}{4}\begin{cases}
	    {2g A_{n}\left( \zeta \right)+1/g}, & n \text{ even} \\
	    {2g B_{n+1}\left( \zeta \right)+1/g}, & n \text{ odd} 
	\end{cases}$ 
\\ [4mm]

BIII\footnotemark[7] & $\displaystyle \frac{\sin^{2}z}{2g^{2}}$ & $\begin{array}{c} \text{$\pi$-antiperiodic} \\ \Psi_{n}(-\pi/2)= -\Psi_{n}(\pi/2) \end{array}$ & $\displaystyle \frac{1}{\sqrt{\pi}}\begin{cases}
	    \text{se}_{n+1}(\zeta,z+\pi/2 ), & n \text{ even} \\
	    \text{ce}_{n}(\zeta,z+\pi/2 ), & n \text{ odd} 
	\end{cases}$ & $\displaystyle \frac{g}{4}\begin{cases}
	    {2g B_{n+1}\left( \zeta \right)+1/g}, & n \text{ even} \\
	    {2g A_{n}\left( \zeta \right)+1/g}, & n \text{ odd} 
	\end{cases}$ 

\end{tabular}

\footnotetext[1]{$H_{n}$ are the Hermite polynomials.}

\footnotetext[2]{$C_{n}^{(\sigma)}$ are the ultraspherical (Gegenbauer) polynomials.}

\footnotetext[3]{the parameters $l>1$ and $j>0$ are defined by $g^{2}=1/l(l-1)$ and $g^{2}=1/j(j+1)$, respectively.}

\footnotetext[4]{$L_{n,l} =  \pi^{-1} e^{-in\pi/2} 2^{-n-l+\frac{1}{2}} \cos(\pi l) \Gamma(-n-l+\tfrac{1}{2}) \sqrt{(n+l) \Gamma(n+1) \Gamma(n+2l)}$}

\footnotetext[5]{$M_{n,j} = \pi^{-1} e^{i(n+1)\pi/2} 2^{-n+j} \Gamma(-n+j+\tfrac{1}{2}) \sqrt{(j-n) \sin(2\pi j) \Gamma(n+1) \Gamma(n-2j)}$; in this case, the potential supports a finite number of energy eigenstates (i.e.~we have $n < j$).}

\footnotetext[6]{$P_{j}^{\sigma}$ denotes the associated Legendre function of degree $j$ and order $\sigma$; and $\{\nu_{n}\}$ represents solutions of equation (\ref{eq:AIII eigenvalue equation}), which must be obtained numerically.  Also, the normalization constants $N_{n,j}^{(\pm)}$ must be obtained numerically.}

\footnotetext[7]{$\ce_{n}$ and $\se_{n}$ are elliptic cosine and sine functions; $A_{n}$ and $B_{n}$ refer to Mathieu characteristic value functions; and $\zeta=1/4g^{2}$.}

}

\end{ruledtabular}

\end{table*}

\subsubsection*{Case AII: $f(x) = 1 - x^{2}$ without cut-off}

The modified commutator is 
\begin{equation}
	[\hat{\phi}_{\k} ,\hat{\pi}_{k} ] = i  \left(1 -  \frac{\hat{\pi}_{\k}^{2}}{M_{\star}} \right),
\end{equation}
and  the eigenvalue equation takes the form
\begin{equation}\label{eq:case AII schrodinger}
	\kappa_{n} \Psi_{n}(z) = \left[ -\frac{1}{2}\frac{\di^{2}}{\di z^{2}} + \frac{\tanh^{2}(z)}{2g^{2}} \right] \Psi_{n}(z),
\end{equation}
with $ \pi_{\k} = M_{\star}^{1/2} \tanh z.$

As in Case AI, this eigenvalue problem has known solutions in terms of ultraspherical polynomials for eigenfunctions satisfying Dirichlet boundary conditions at infinity
\begin{equation}
	\lim_{z\rightarrow \pm \infty} \Psi_{n}(z) = 0,
\end{equation}
and inner product  $(\Psi_{n},\Psi_{m})_{\infty} = \delta_{m,n}$.  (Note that in this case, the square-integrability of the wavefunction admits no other boundary conditions.)  Explicit formulae for the eigenfunctions and energy eigenvalues are given in Table \ref{tab:eigenfunctions}, and  are plotted in Figures \ref{fig:eigenfunctions} and \ref{fig:eigenvalues}, respectively. 

Unlike the previous cases considered, the potential in the Schrodinger equation (\ref{eq:case AII schrodinger}) only supports a finite number of bound states, which do not form a complete energy eigenfunction basis (unless the potential is put in a box). The is a direct consequence of the finite height of the potential appearing in (\ref{eq:case AII schrodinger}), as can be seen in Figure \ref{fig:eigenfunctions}.  

\subsubsection*{Case AIII: $f(x) = 1 - x^{2}$ with cut-off}

Since the energy eigenfunction basis of Case AII is not complete, the formulae of \S\ref{sec:stat mech} are not applicable and we cannot directly obtain the blackbody spectrum.  However, we can recover a complete eigenfunction basis if we impose Dirichlet boundary conditions on a finite boundary, i.e. 
\begin{equation}\label{eq:AIII BC}
	\Psi_{n}(\pm \ell) = 0.
\end{equation}
where $\ell$ is an adjustable dimensionless (box) parameter.  In this case  the exact solution of (\ref{eq:case AII schrodinger}) may be written in terms of associated Legendre functions as   in Table \ref{tab:eigenfunctions}.  It can be shown that the energy eigenvalues $ \epsilon_{n}$are related to the roots $\nu_{n}$ of the equation
\begin{equation}\label{eq:AIII eigenvalue equation}
	[P_{j}^{i\nu_{n}}(\tanh\ell)]^{2} = [P_{j}^{i\nu_{n}}(-\tanh\ell)]^{2}, 
\end{equation}
via the relation 
\begin{equation}
         \epsilon_{n} = \frac{1}{2} (g^{2} \nu_{n} + 1).
\end{equation}
Note that $\nu_{n}$ can be either real or imaginary, corresponding to eigenmodes with energy greater or less than $\M$, respectively.  Unfortunately, (\ref{eq:AIII eigenvalue equation}) is not exactly solvable for $\nu_n$, but it is possible to obtain the energies numerically for a given value of $\ell$.  Examples of numerically obtained eigenfunctions and eigenvalues are shown in Figures \ref{fig:eigenfunctions} and \ref{fig:eigenvalues} for $\ell = 5$.

Before we move one, it may be useful to interpret the boundary condition (\ref{eq:AIII BC}) in terms of the original field variables.  It is easy to see that the boundary condition implies
\begin{equation}
	|\pi_{\k}| \le \M^{1/2}\tanh\ell.
\end{equation}
Hence by enforcing (\ref{eq:AIII BC}), we are effectively imposing a cut-off for the field momentum $\pi_{\k}$.  We will see below that this cutoff is necessary to tame UV divergences in quantities such as the internal energy of the scalar gas at finite temperature.

\subsection{Polymer quantization}\label{sec:polymer}

As described elsewhere \cite{Ashtekar:2002sn,Seahra:2012un}, there is no explicit realization of the momentum operator $\hat{\pi}_{\k}$ in polymer quantum mechanics.   However, we can write an ``effective'' momentum operator as 
\begin{equation}
	  \hat{\pi}^{\star}_{\k} =  \frac{\hat{U}_{\lambda_{\star}}-\hat{U}_{\lambda_{\star}}^{\dag}}{2i\lambda_{\star}}.
\end{equation} 
where $\lambda_{\star} \equiv M_{\star}^{-1/2}$ is a fixed parameter with dimensions of $(\text{mass})^{-1/2}$, and $\hat{U}_{\lambda}$ is an operator which induces translations of magnitude $\lambda$ in the field amplitude $\phi_{\k}$:
\begin{equation}
	[\hat\phi_{\k},\hat{U}_{\lambda}] = -\lambda\hat{U}_{\lambda}, \quad \hat{U}_{\lambda}^{\dag}  \hat\phi_{\k} \hat{U}_{\lambda} = \hat{\phi}_{\k} - \lambda \hat{\mathbf{1}}.
\end{equation}
We now introduce a basis $|\pi_{\k}\rangle$ and express the Hamiltonian eigenstates $|n_{\k}\rangle$ as wavefunctions:
\begin{equation}\label{eq:momentum wavefunction}
	\langle \pi_{\k} | n_{\k} \rangle = \psi_{n}(\pi_{\k}).
\end{equation}
The action of $\hat{\phi}_{\k}$ and $\hat{U}_{\k}$ on these wavefunctions is 
\begin{subequations}
\begin{align}
	\langle \pi_{\k} | \hat\phi_{\k} | n_{\k} \rangle &  = i {\di_{\pi_{\k}}} \psi_{n}(\pi_{\k}), \\ \langle \pi_{\k} | \hat{U}_{\lambda} |n_{\k} \rangle & = \exp ( i\lambda \pi_{\k} )\psi_{n}(\pi_{\k}).
\end{align}
\end{subequations}
The energy eigenvalue equation (\ref{eq:eigenvalue problem}) becomes 
\begin{equation}\label{eq:polymer H}
	E_{n,k} \psi_{n}(\pi_{\k}) = \left[ \frac{\sin^{2} (\lambda_{\star}\pi_{\k})}{2\lambda_{\star}^{2}} - \frac{k^{2}}{2} \frac{\di^{2}}{\di\pi_{\k}^{2}} \right] \psi_{n}(\pi_{\k}),
\end{equation}
The   inner product is 
\begin{equation}
	   \int_{-\pi/2\lambda_{\star}}^{\pi/2\lambda_{\star}} \psi_{m}^*(\pi_{\k})  \psi_{n}(\pi_{\k})    d\pi_{\k} = \delta_{m,n}.
\end{equation}

To solve the eigenvalue problem, it is convenient to transform to dimensionless quantities:
\begin{align}\label{eq:polymer defs}
	\nonumber g & = \lambda^{2}_{\star} k =  k M_{\star}^{{-1}}, &  \pi_{k} & = \sqrt{\M} z, \\ \psi_{n}(\pi_{\k}) & = \M^{-1/4} \Psi_{n}(z), & E_{n,k} & = g^{2} \M \kappa_{n}.
\end{align}
In terms of these, the eigenvalue problem becomes
\begin{equation}\label{eq:polymer H simple}
	\kappa_{n} \Psi_{n}(z) = \left[  - \frac{1}{2} \frac{\di^{2}}{\di z^{2}} + V(z) \right] \Psi_{n}(z),
\end{equation}
where the potential is
\begin{equation}
	V(z) = \frac{\sin^{2}z}{2g^{2}}.
\end{equation}
We see that the dimensionless quantities and Schr\"odinger equation arising from this analysis are analogous to the definitions (\ref{eq:case A defs}) and the ODE (\ref{eq:case A Schrodinger}); in both we have a similar quantum mechanics problem.
 
This differential equation is simply related to the well-known Mathieu equation.   A complete basis of eigenfunctions of definite periodicity can be written down in terms of elliptic cosine $\text{ce}_{n}$ and elliptic sine $\text{se}_{n}$ functions, while the energy eigenvalues are given by the Mathieu characteristic value functions $A_{n}$ and $B_{n}$ \cite{abramowitz+stegun}.  

To completely specify the eigenvalue problem, we must impose boundary conditions on $\Psi_{n}$ at $z = \pm \pi/2$.  The details of the polymer construction lead to the following condition:
\begin{equation}
	\Psi_{n}(-\pi/2) = e^{i\alpha} \Psi_{n}(\pi/2),
\end{equation}
where $\alpha$ is a freely-specifiable phase angle related to the lattice offset one uses to specify a super-selected Hilbert space.  We consider two different choices: $\alpha= 0$ or $\pi$-periodic (case BII) and $\alpha=\pi$ or $\pi$-antiperiodic (case BIII).  In addition, we consider the (logically distinct) Dirichlet boundary conditions $\Psi_{n}(-\pi/2)=\Psi_{n}(\pi/2)=0$.  This last choice (case BI) is motivated in Ref.~\cite{Seahra:2012un},  where such a boundary condition was introduced in the context of polymer quantized fluctuations during inflation.  The three resulting families of eigenfunctions and eigenvalues are given in Table \ref{tab:eigenfunctions} and plotted in Figures \ref{fig:eigenfunctions} and \ref{fig:eigenvalues}, respectively.  All three classes of boundary condition recover the usual energy levels of the simple harmonic oscillator in the $g \rightarrow 0$ limit,  
\begin{equation}
	\varepsilon_{n} \approx g(n + 1/2), \quad g \ll 1.
\end{equation}
However at high $g$, the eigenvalues are sensitive to the boundary conditions; in particular, the energy differences obey
\begin{equation}
	\Delta\varepsilon_{n} \approx g^{2}\begin{cases}
	\frac{1}{2} n(n+2) , & \text{case BI}, \\
	 2 \left( \lfloor \frac{n+1}{2} \rfloor \right)^{2} , & \text{case BII}, \\
	\lceil \frac{n-1}{2} \rceil \lceil \frac{n+1}{2} \rceil + \frac{1}{4g^{2}} \delta_{n,1} , & \text{case BIII}.
	\end{cases}
\end{equation}
Here $\lceil \cdots \rceil$ and $\lfloor \cdots \rfloor$ are the ceiling and floor functions, respectively.  We note that  $\Delta\varepsilon_{n} \approx g^{2} f_{n}$ for cases BI and BII, where the numbers $\{ f_{n} \}_{n=0}^{\infty}$ do not depend on $g$.

\section{Numeric calculations of blackbody spectra}\label{sec:numeric}

\begin{figure*}
	\includegraphics[width=\textwidth]{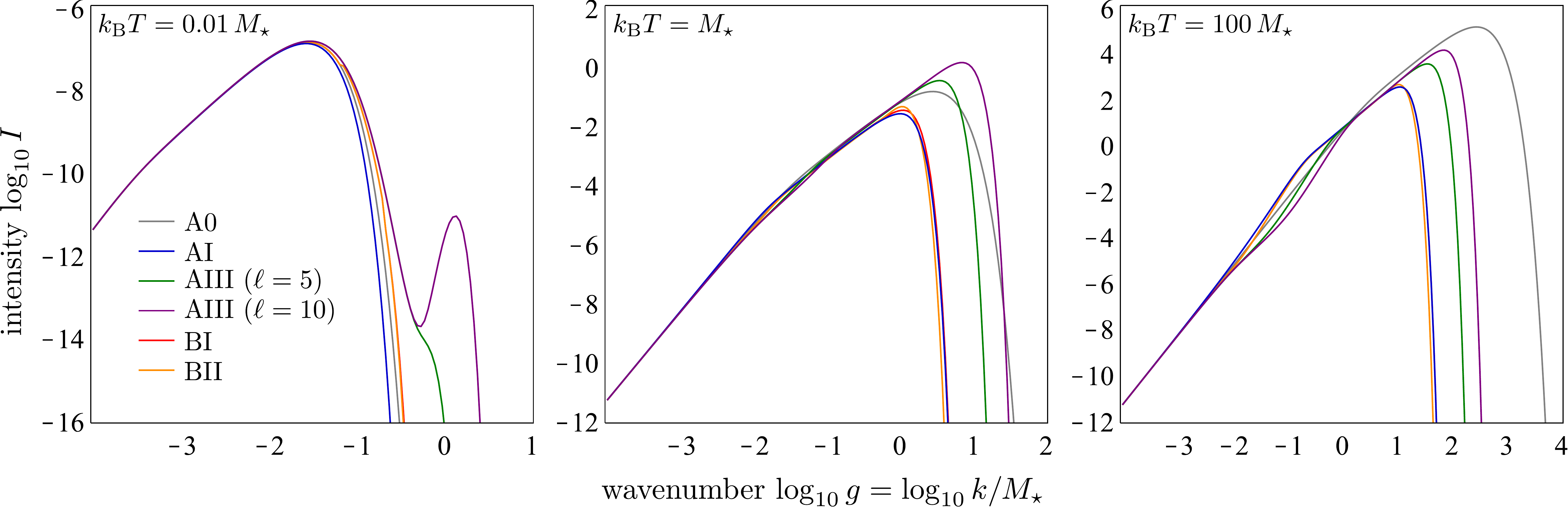}
	\caption{Blackbody radiation spectra for various scalar field gases at low (\emph{left}), moderate (\emph{centre}) and high (\emph{right}) temperature.  In all situations, the spectra match the standard A0 result at small wavenumbers $k/M_{\star} \ll 1$.  At low temperature there is close agreement near the peak of the intensity distributions, which implies the thermodynamic properties of the system are virtually indistinguishable.  At moderate and high temperatures the peak degeneracy is lifted to some degree, although cases A1, BI and BII remain very close together.  At high temperature, the intensity of the standard case A0 is several orders of magnitude greater than the others for high $g$.  This reflects the fact  energy differences scale as $g^{2}$ for the modified cases; i.e., they are larger than in the A0 case.}\label{fig:intensity}
\end{figure*} 
\begin{figure}
	\includegraphics[width=0.8\columnwidth]{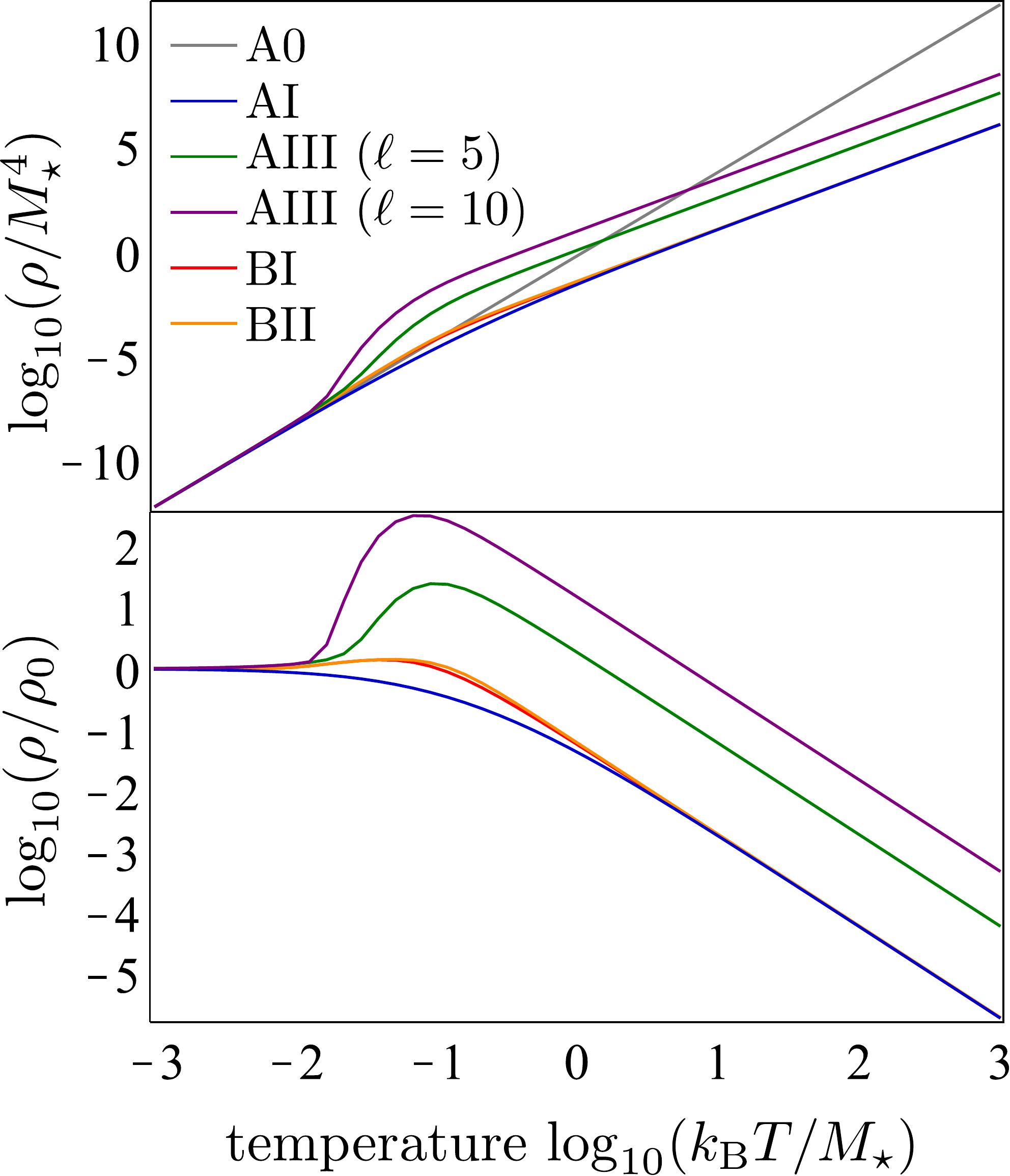}
	\caption{Stefan-Boltzmann law for various scalar field gases with modified oscillator spectra.  Here, the energy density is defined by $\rho = \mathcal{U}/V$, where $\mathcal{U}$ is the total energy and $V$ is the volume.  In the bottom panel, we show the ratio of the modified energy densities to the conventional (case A0) result $\rho_{0} =\pi^{2} k_\text{B}^{4}T^{4}/15$.}\label{fig:U}
\end{figure}     
\begin{figure}
	\includegraphics[width=\columnwidth]{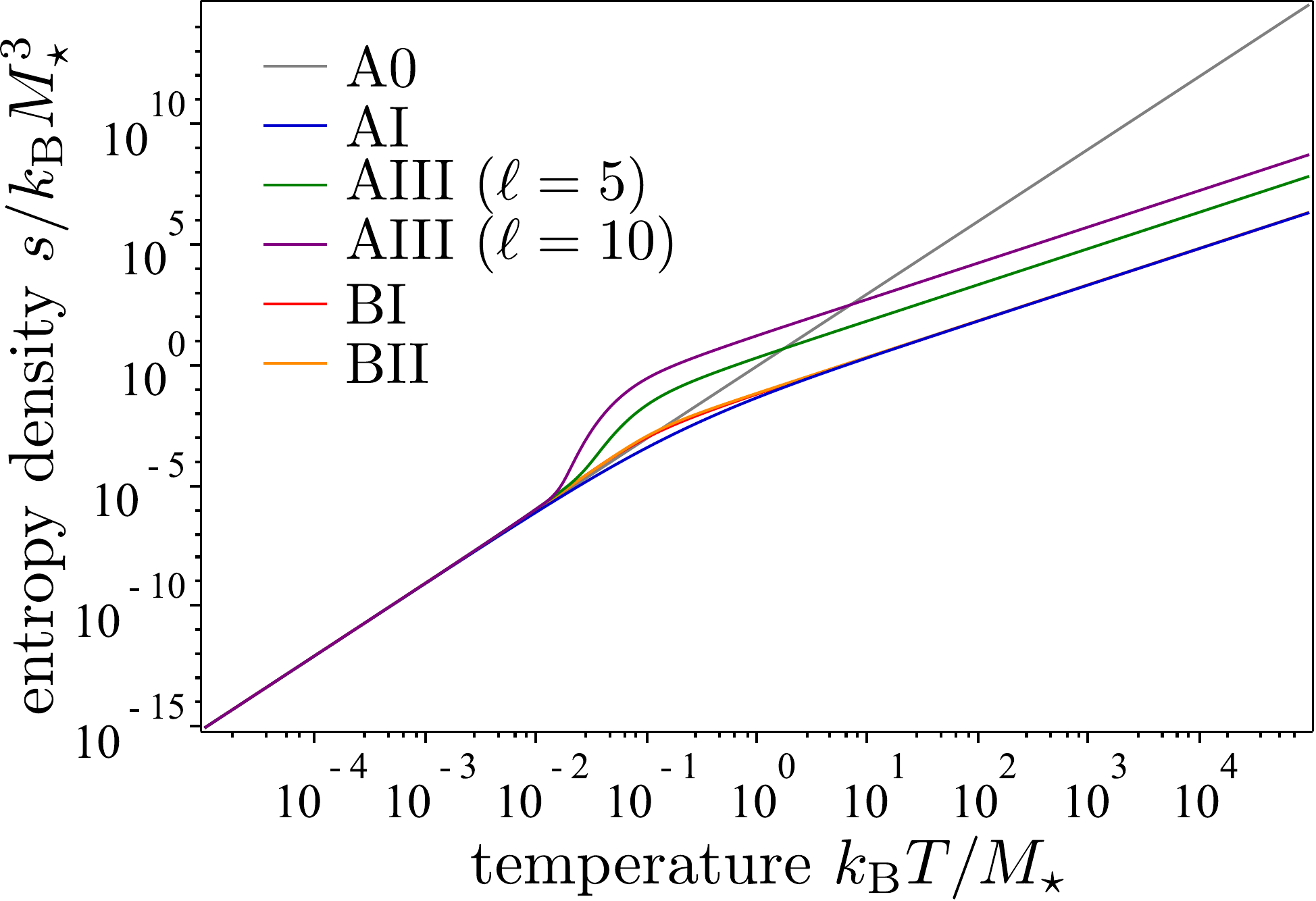}
	\caption{Entropy density for various scalar field gases with modified oscillator spectra}\label{fig:entropy}
\end{figure} 
\begin{figure}
	\includegraphics[width=\columnwidth]{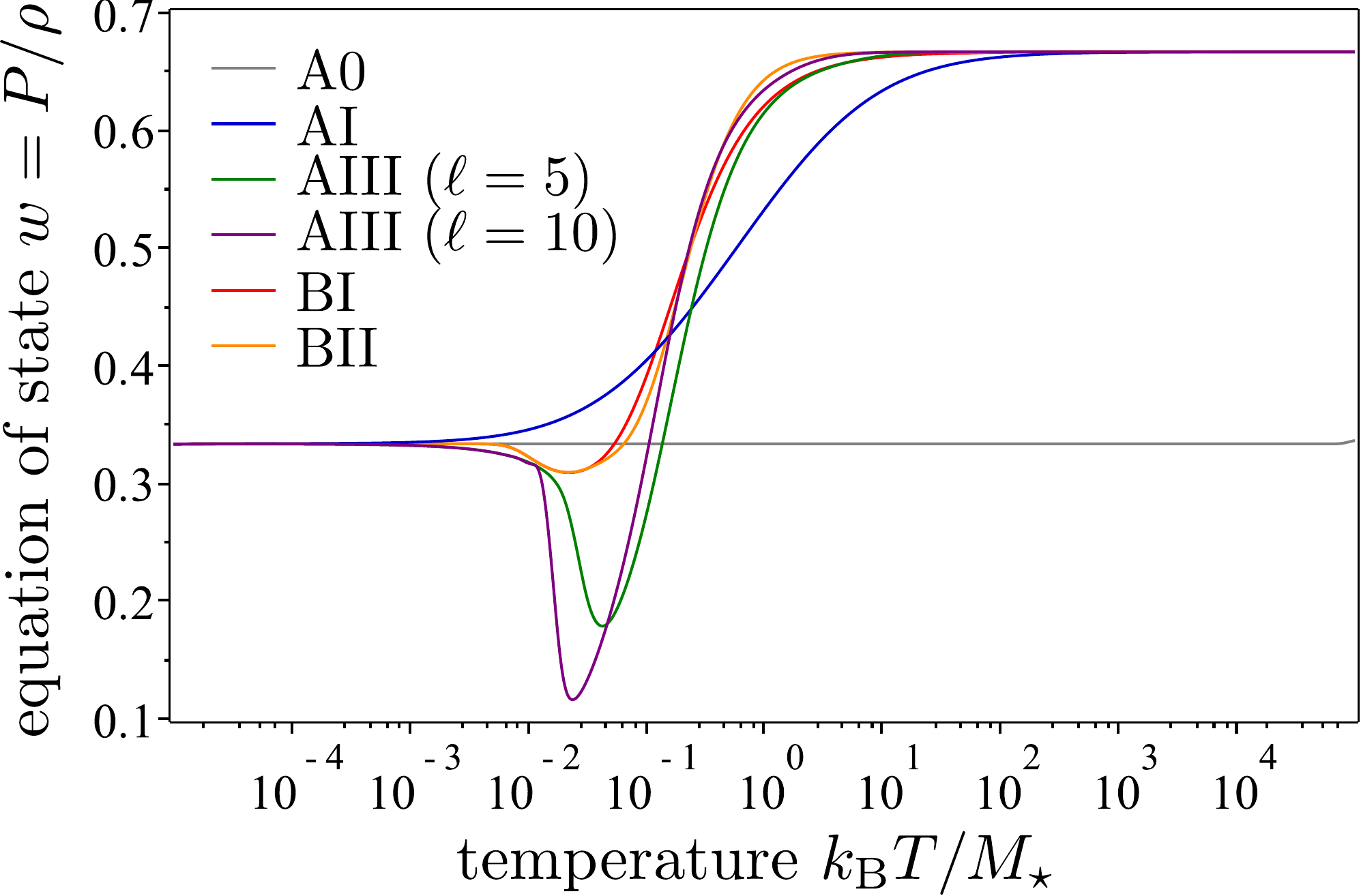}
	\caption{Equation of state for various scalar field gases with modified oscillator spectra.  As expected from the general discussion of \S\ref{sec:stat mech}, the equation of state interpolates between that of ordinary radiation $w=1/3$ at low temperature and $w = 2/3$ at high temperature.}\label{fig:eos}
\end{figure}        
In this section, we analyze the thermal properties of scalar field gases governed by the modified oscillator spectra discussed in the previous section.

\subsection{Cases AI, AIII, BI and BII}

These scenarios are characterized by a complete basis of energy eigenfunctions. The energy differences satisfy
\begin{equation}
	\Delta\varepsilon_{n} \approx \begin{cases} gn, & g \ll 1, \\ g^2 f_{n}, & g \gg 1. \end{cases}
\end{equation} 
where the numbers $f_{n}$ are independent of $g$.  According to the general discussion of \S\ref{sec:low and high T}, we expect the total energy  to obey
\begin{equation}
	\mathcal{U} \propto \begin{cases} T^{4}, & k_\text{B}T \ll \M, \\ T^{5/2}, & k_\text{B}T \gg \M. \end{cases}
\end{equation}
 This result is confirmed  numerically by calculating the dimensionless intensity (\ref{eq:intensity}) using (\ref{eq:rho}), and then integrating to get $\mathcal{U}$.  We approximate the infinite sum in equation (\ref{eq:intensity}) by truncating at some large value of $n$.  The cutoff is determined by demanding that the fractional change in the sum induced by an additional term is than small tolerance (which we take to be $\sim 10^{-6}$).

Figure \ref{fig:intensity} shows  the intensity of blackbody radiation at low, moderate and high temperatures for several  deformed oscillator spectra; for comparison  the standard result   for the conventional oscillator spectrum is case A0.  Figure \ref{fig:U} shows the internal energy density $\rho$ as a function of temperature; i.e., the Stefan-Boltzmann Law.  We see that the usual $\mathcal{U} \propto T^{4}$ behaviour is recovered at low temperature, whereas  at high temperature it matches  the analytic expectation  $\mathcal{U} \propto T^{5/2}$. 

From the numerically obtained energy density, Eqns. (\ref{eq:entropy}) and (\ref{eq:pressure}) give the entropy density and effective equation of state $w = P/\rho$ as functions of temperature.  These are shown in Figures \ref{fig:entropy} and \ref{fig:eos}, respectively;  we see   that the equation of state  interpolates between $P= \frac{1}{3} \rho$ at low $T$ and $P = \frac{2}{3} \rho$ at high $T$, consistent with the low and high temperature approximations in  \S\ref{sec:low and high T}.

\subsection{Case AII}

This case involves a Schr\"odinger equation with potential $\propto \tanh^{2}z$ and Dirichlet boundary conditions at infinity.  We can view this scenario as the $\ell \rightarrow \infty$ limit of case AIII.  In this case, we can use the Bohr-Sommerfeld quantization condition to estimate the energy eigenvalues of the excited ($n \ge 1$) states:
\begin{equation}
	n\pi = \int_{-\ell}^{\ell} \sqrt{ \kappa_{n} - V(z) } dz, \quad V(z) = \frac{\tanh^{2}z}{2g^{2}}.
\end{equation}
In the $\ell \rightarrow \infty$ limit, the integral on the right will be dominated by contributions from $|z| \gg 1$.  As a first approximation, one can just neglect the potential in this limit yield the energy eigenvalues of a particle in a box:
\begin{equation}
	E_{n,k} \approx \frac{\M}{2} + \frac{k^{2}\pi^{2}n^{2}}{4\M\ell^{2}}, \quad n \ge 1.
\end{equation}
For the ground state,  the exact result from Table \ref{tab:eigenfunctions} is
\begin{equation}
	E_{0,k} = \frac{k}{4} \left( \sqrt{4+\frac{k^{2}}{\M}} - \frac{k}{\M} \right).
\end{equation}
For $\ell \rightarrow \infty$ the energy levels are closely spaced so sums over $n$ are well approximated by integrals.  Substituting these energy eigenvalues into (\ref{eq:normal partition}) and (\ref{eq:normal energy}) yield the average energy in a $k$ mode is
\begin{equation}
	\bar{E}_{\k}(\beta) = \frac{1}{2\beta} \left[ 1 + \mathcal{O}\left(\frac{\beta \M^{3}}{k^{2}} \right) \right].
\end{equation}
For high $k$, this is the familiar ``particle in a box''  result that the average energy is a function of temperature only.  Since $\bar{E}_{\k}(\beta)$ approaches a constant, the integral for the internal energy (\ref{eq:normal total energy}) is UV divergent.  It would seem that this particular class of modification does not lead to a viable finite temperature thermal system.  The divergence of the internal energy as $\ell \rightarrow \infty$ seems to be consistent with the finite $\ell$ curves in Figure \ref{fig:U}, where  $\rho$ for $\ell = 10$ is about an order of magnitude larger than for $\ell = 5$ at high temperature.

\subsection{Case BIII}

This case differs from the other polymer quantization cases (BI and BII) in that the energy differences satisfy
\begin{equation}
\Delta \varepsilon_{1} \approx \frac{1}{4}, \quad \Delta\varepsilon_{n} \propto g^{2}, \quad g \gg 1, \quad n \ge 2.
\end{equation}
That is, $\Delta\varepsilon_{1}$ remains finite in the $g \rightarrow \infty$ limit.  From this, it follows that at small wavelengths the intensity (\ref{eq:intensity}) satisfies 
\begin{equation}
	I \propto g^{2}, \quad \tilde\beta g^{2} \gg 1.
\end{equation}
This in turn implies that the integral for the internal energy (\ref{eq:rho}) isUV divergent.  As for Case AII, it would seem that this particular class of modification does not lead to a viable finite temperature thermal system.

\section{Discussion}\label{sec:discussion}

We applied alternative quantization methods that come with a fundamental scale to the thermodynamics of  a scalar gas. The various quantization prescriptions change the microscopic energy levels through potentials and boundary conditions (as shown in Fig. \ref{fig:eigenfunctions}), which in turn impact the thermodynamics.    
  
 We obtained  two classes of results, one where the internal energy of the gas is divergent,  and the other where it is finite. In the latter case {\it the UV behaviour is the same for all the quantization prescriptions considered}. 
 
 The two cases in the first category are the polymer quantized scalar field with $\pi$-anti periodic boundary conditions (case BIII),  and   where the commutator of  Fourier space pause space variables is modified to  
\begin{equation}\label{eq:mod commutator 1}
	[\hat{\phi}_{\k} ,\hat{\pi}_{\k} ] = i  \left(1 -  \frac{\hat{\pi}_{\k}^{2}}{M_{\star}} \right),
\end{equation}
with no additional bounds on the expectation value of $\hat{\pi}_{\k}^{2}$ (case AII).  

All the other cases  considered---which include (\ref{eq:mod commutator 1}) with a momentum cutoff $|\pi_{\k}| < \M$---share the property that the energy levels of  Fourier oscillators scale as  $k^{2}/\M$ for $k \gg \M$.  This implies that at high temperature, the internal energy is proportional to $T^{5/2}$ and the equation of state is $P= 2\rho/3$.  

The reason for this generic high temperature behaviour stems from the nature of the modified oscillator potentials governing each of the Fourier modes (cf.\ Figure \ref{fig:eigenfunctions}).  For $k \gg \M$, these potentials and boundary conditions all reproduce the energy levels  of a non-relativistic particle in a box, i.e. $E_{n.k} \propto k^2$ (modulo some degeneracies in the polymer case).   This  is the key feature that leads to the same high $T$ characteristics.  We expect that any modified quantization that leads to a potential that looks like a square well for $k \gg \M$ will give similar results. 

For  blackbody radiation in $d$ spatial dimensions,   the Stefan-Boltzmann law is $\mathcal{U} \propto T^{1+d}$.  Thus  one can use the temperature scaling of the internal energy as a measure of the effective dimensionality of space.  In our case, this gives a curious fractional effective space dimension $d=3/2$  (or  spacetime dimension 5/2) at high temperature.

It is interesting in this context to note that  various approaches to  quantum gravity give indications that the effective spacetime dimension at high energy is closer to two than four. These results are summarized in \cite{Carlip-2d}, and some cosmological consequences are discussed in \cite{Mureika:2011bv,Mureika:2012na}.  Could it be that there is dimensional reduction in quantum field theory in the UV with alternative quantizations?  Such a scenario is not envisaged in conventional effective field theory, where the dimension of spacetime is fixed at the outset at any energy scale. 

It may be interesting to couple the modified scalar fields discussed in this paper to Friedmann-Roberstson-Walker cosmologies to see if the gradual ``phase change'' of the equation of state from $w = 1/3$ to $w=2/3$ has an impact on the high energy radiation phase of the universe's expansion.  It is known that big bang nucleosynthesis requires that the universe have  equation of state $P=\rho/3$ at temperature of $T \sim 1\,\text{MeV}$.  This places a rather weak constraint of $\M \gtrsim 1\,\text{MeV}$ on the new physics energy scale \footnote{Other physics (such as from the Large Hadron Collider and inflation) can certainly place more stringent bounds on $\M$.}.  

These are among the  many  interesting questions concerning the physical implications of  quantizations prescriptions that come with both $\hbar$ and a mass scale $\M$. 

\begin{acknowledgments}

We would like to thank Dawood Kothwala for many useful discussions.  We are funded by NSERC of Canada.

\end{acknowledgments}

\bibliography{blackbody}

\end{document}